\documentclass[10pt,twocolumn,letterpaper]{article}

\usepackage{iccv}
\usepackage{times}
\usepackage{epsfig}
\usepackage{graphicx}
\usepackage{amsmath}
\usepackage{amssymb}
\usepackage{authblk}
\usepackage{booktabs}
\usepackage{comment}
\renewcommand\Affilfont{\fontsize{7.7}{14.4}\selectfont}

\newcommand\blfootnote[1]{%
  \begingroup
  \renewcommand\thefootnote{}\footnote{#1}%
  \addtocounter{footnote}{-1}%
  \endgroup
}
\usepackage[hang,flushmargin]{footmisc}
\makeatletter
\renewcommand\AB@affilsepx{  \hspace{1 mm}  \protect\Affilfont}
\makeatother

\usepackage[breaklinks=true,bookmarks=false]{hyperref}

\iccvfinalcopy 

\def\iccvPaperID{****} 

\ificcvfinal\fi

\begin{document}

\title{AIM 2019 Challenge on Image Demoireing: Methods and Results}

\author{
Shanxin Yuan \hspace*{4mm}
Radu Timofte \hspace*{4mm}
Gregory Slabaugh \hspace*{4mm}
Ale\v{s} Leonardis  \\
Bolun Zheng \hspace*{4mm}
Xin Ye \hspace*{4mm} 
Xiang Tian \hspace*{4mm}
Yaowu Chen \hspace*{4mm}
Xi Cheng \hspace*{4mm}
Zhenyong Fu \\
Jian Yang \hspace*{4mm}
Ming Hong \hspace*{4mm} 
Wenying Lin \hspace*{4mm} 
Wenjin Yang \hspace*{3.5mm}
Yanyun Qu \hspace{3.5mm}
Hong-Kyu Shin \\
Joon-Yeon Kim \hspace*{3.5mm} 
Sung-Jea Ko \hspace*{3.5mm} 
Hang Dong \hspace{3.5mm}
Yu Guo \hspace{3.5mm}
Jie Wang \hspace{3.5mm}
Xuan Ding \hspace{3.5mm}
Zongyan Han \\
Sourya Dipta Das\hspace*{4mm}
Kuldeep Purohit \hspace*{4mm}
Praveen Kandula \hspace*{4mm}
Maitreya Suin \hspace*{4mm}
A. N. Rajagopalan \hspace*{4mm}
}
\affil[]{}

\maketitle
\ificcvfinal\thispagestyle{empty}\fi

\blfootnote{
S. Yuan (shanxin.yuan@huawei.com, Huawei Noah's Ark Lab), R.
Timofte, G. Slabaugh and A. Leonardis are the AIM 2019 challenge organizers, while the other authors participated in the challenge.\\
Appendix A contains the authors’ teams and affiliations.\\
AIM webpage: \url{http://www.vision.ee.ethz.ch/aim19/}
}

\begin{abstract}
This paper reviews the first-ever image demoireing challenge that was part of the Advances in Image Manipulation (AIM) workshop, held in conjunction with ICCV 2019.  This paper describes the challenge, and focuses on the proposed solutions and their results. Demoireing is a difficult task of removing moire patterns from an image to reveal an underlying clean image. 
A new dataset, called \emph{LCDMoire} was created for this challenge, and consists of 10,200 synthetically generated image pairs (moire and clean ground truth).
The challenge was divided into 2 tracks. 
Track 1 targeted fidelity, measuring the ability of demoire methods to obtain a moire-free image compared with the ground truth,
while Track 2 examined the perceptual quality of demoire methods.
The tracks had 60 and 39 registered participants, respectively. A total of eight teams competed in the final testing phase. The entries span the current the state-of-the-art in the image demoireing problem.
\end{abstract}

\section{Introduction}
\label{report:into}

It is becoming increasingly common for people to take pictures of screens, in particular to quickly save useful information for later reference. 
Indeed, screens are ubiquitous in our daily lives---they can be found at home, in offices, public places such as transportation stations, educational institutions etc.  
For example, when attending a scholarly conference one may wish to take pictures of slides displayed on a screen, and study them carefully later.
%
Moire patterns appear when two repetitive patterns interfere with each other. In the case of digital photography of screens, moire patterns occur when the screen's subpixel layout interferes with the camera's color filter array (CFA).
%
%
Digital image quality has been improving over the past years, with great improvements introduced in image denoising~\cite{DnCNN}, image demosaicing~\cite{DemosaicNet}, sharpening \cite{romano2016raisr}, automatic white balancing~\cite{FFCC}, and high dynamic range compression~\cite{hdrnet}. However, current image signal processing (ISP) pipelines often produce strong moire patterns when presented a photograph of a screen taken with a digital camera.

Given the increasing occurrence of moire images resulting from digital photography of screens, coupled with the comparatively less attention the demoireing problem has received in the the community \cite{liu2018demoir, sun2018moire}, this demoireing challenge was proposed to fill this gap. By engaging the academic community with this challenging image enhancement problem, a variety of methods have be proposed, evaluated, and compared using a common dataset.

\section{The Challenge}
\label{report:challenge}

The demoireing challenge was hosted jointly with the first \textit{Advances in Image Manipulation (AIM 2019)} workshop held in conjunction with the International Computer Vision Conference (ICCV) 2019, Seoul, Korea.
The task of demoireing was to create an algorithm to remove the moire effect from an input image degraded by moire effect.  To help build algorithms, particularly those based on machine learning, a set of example images with and without the moire effect was produced.

\subsection{Dataset}

As an essential step towards reducing moire effects, we proposed a novel synthetic dataset, namely LCDMoire~\cite{AIM19demoireDataset}.
It consists of 10,000 training, 100 validation, and 100 test images with 1024 $\times$ 1024 resolution.
The images are high quality in terms of the reference frames, different moire patterns, noise, and compression artifacts. It also covers balanced content of images containing figures and text.
The clean images are extracted from existing conference papers (\eg, papers from ICCV, ECCV, and CVPR proceedings). The dataset consists of images of figures and text in equal proportion. 
The moire images are generated through a pipeline, similar to~\cite{liu2018demoir}, modeling the picture taking process when a smartphone is used to take pictures from an LCD screen displaying a clean image. Post-processing was conducted on the distributions of moire patterns and ensured the same power spectra distribution among the three subsets.

\subsection{Tracks and Evaluation}
The challenge had two tracks: \textbf{Fidelity} and \textbf{Perceptual}.

\textbf{Track 1 Fidelity:} In this track, participants developed demoire methods to achieve a moire-free image with the best fidelity compared to the ground truth moire-free image. We used the standard Peak Signal To Noise Ratio (PSNR) and, additionally, the Structural Similarity (SSIM) index as often employed in the literature. Implementations are found in most of the image processing toolboxes. For each dataset we report the average results over all the processed images.

\textbf{Track 2 Perceptual:} In this track, participants developed demoire methods to achieve a moire-free image with the best perceptual quality (measured by Mean Opinion Score (MOS)) when compared to the ground truth moire-free image. In the Mean Opinion Score, human subjects are invited to express their opinions comparing the demoired image to the clean image. Along with the MOS results, we also report the standard Peak Signal To Noise Ratio (PSNR) and the Structural Similarity (SSIM) as a reference. However, the MOS is the ranking metric in this track.

\subsection{Competition}

\textbf{Platform:} The CodaLab platform was used for this competition. To access the data and submit their demoired image results to the CodaLab evaluation server each participant had to register.

\textbf{Challenge phases:} 
(1) Development (training) phase: the participants received both moire and moire-free training images of the LCDMoire dataset~\cite{AIM19demoireDataset}; 
(2) Validation phase: the participants had the opportunity to test their solutions on the moire validation images and to receive immediate feedback by uploading their results to the server. A validation leaderboard was available; 
(3) Final evaluation phase: after the participants received the moire test images and clean validation images, they had to submit both their demoired images and a description of their method before the challenge deadline. One week later the final results were made available to the participants.

\section{Results}
\label{report:results}

\begin{table*}[th]
\normalsize 
  \centering
  \resizebox{2.\columnwidth}{!}{
  \begin{tabular}{lccccccc}
  \toprule 
  \bf Team & \bf User name & \bf PSNR  & \bf SSIM & \bf Time & \bf GPU & \bf Platform & \bf Loss \\ 
  \midrule 

Islab-zju
& \emph{q19911124}
& 44.07
& 0.99 
& 0.25
& 2080Ti
& Tensorflow 
& L1 + L1 sobel
\\

MoePhoto \cite{cheng2019multi}
& \emph{opteroncx}
&41.84
& 0.98
& 0.10
& Titan V
& Pytorch 
& L1 charbonnier loss
\\

KU-CVIP
& \emph{GibsonGirl}
& 39.54
& 0.97
& 0.07
& Titan X Pascal
& Pytorch 
& L1
\\

XMU-VIPLab
&\emph{xdhm2017}
& 39.35 
& 0.98
& 0.13
& RTX 2080
& Pytorch
& L1 + perceptual loss + angular loss
\\

IAIR
& \emph{wangjie}
& 38.54 
& 0.98
& 0.43
& 1080Ti 
& Pytorch 
& L2 + perceptual loss
\\

PCALab
&\emph{Hanzy}
& 33.72
& 0.98
& 0.02
& 2080Ti 
& Pytorch 
& L1 + contextual loss
\\

IPCV\_IITM
&\emph{kuldeeppurohit3}
& 32.23
& 0.96
& -
& Titan X 
& Pytorch
& -
\\

Neuro
&\emph{souryadipta}
& 29.87
& 0.92
& 0.25
& 1080Ti
& Pytorch
& -
\\

  \bottomrule
  \end{tabular}}
  
    \caption{Results and rankings of methods submitted to the \textbf{fidelity track}. The methods are ranked by PSNR values, \textit{Time} is for run time per image and measured in seconds.}
  \label{tab:fidelity} 

\end{table*}

\begin{table*}[ht]
\normalsize 
  \centering
  \resizebox{2.\columnwidth}{!}{
  \begin{tabular}{lccccccc}
  \toprule 
  \bf Team & \bf User name & \bf PSNR  & \bf SSIM & \bf Time &\bf GPU  & \bf MOS & \bf Loss \\ 
  \midrule 

Islab-zju
& \emph{q19911124}
& 44.07
& 0.99 
& 0.25
& 2080Ti
& 241
& L1 + L1 sobel
\\

XMU-VIPLab
&\emph{xdhm2017}
& 39.35 
& 0.98
& 0.13
& RTX 2080
& 216
& L1 + perceptual loss + angular loss
\\

MoePhoto \cite{cheng2019multi}
&\emph{opteroncx}
& 31.07
& 0.96
& 0.11
& Titan V
& 215
& L1 + perceptual loss
\\

KU-CVIP
& \emph{GibsonGirl}
& 39.54
& 0.97
& 0.07
& Titan X Pascal
& 210
& L1
\\

Neuro
&\emph{souryadipta}
& 29.87
& 0.92
& 0.25
& 1080Ti
& 187
& -
\\

PCALab
&\emph{Hanzy}
& 30.48
& 0.96
& 0.01
& 2080Ti 
& 183
& L1 loss
\\

IPCV\_IITM
&\emph{kuldeeppurohit3}
& 32.23
& 0.96
& -
& Titan X 
& 169
& -
\\

  \bottomrule
  \end{tabular}}
  
    \caption{Results and rankings of methods submitted to the \textbf{perceptual} track. The methods are ranked by MOS scores, \textit{Time} is for run time per image and measured in seconds.}
  \label{tab:perceptual} 

\end{table*}

From 60 registered participants, eight teams entered in the final phase and submitted results, codes/executables, and factsheets. Table~\ref{tab:fidelity} and Table~\ref{tab:perceptual} report the final test results, rankings of the challenge, self-reported runtimes and major details from the factsheets. 

For the \textbf{fidelity} track, the leading entry as shown in Table~\ref{tab:fidelity} was from the Islab-zju team, scoring a PSNR of $44.07$ dB.  Second and third place entries were by teams MoePhoto and KU-CVIP respectively.

For the \textbf{perceptual} track, the leading entry as shown in Table~\ref{tab:perceptual} was also from the Islab-zju team, scoring a MOS of $241$.  Second and third place entries were by teams XMU-VIPLab and MoePhoto respectively.  

Looking across entries, some interesting trends were noted.  In particular,
\begin{itemize}
\item \textbf{Ensembles:} Some solutions used a self-ensemble~\cite{SevenWays} that averages
the results from flipped and rotated inputs at test time. After going through a demoire approach, 
the inverse flipping/rotation was applied and the ensemble of demoired images were averaged.

\item \textbf{Augmentation:} Training data augmentation strategy such as flips and rotations by 90 degrees~\cite{SevenWays} were
employed by many of the participants. 

\item \textbf{Multi-scale strategy:} Most solutions adopted a multi-scale strategy as a mechanism to handle moire patterns of different frequencies.

\item \textbf{Similar, or different losses between tracks}: Although many solutions choose the same loss function for the two tracks, a few tried different losses for each track, trying to emphasize reconstruction for the fidelity track and perceptual quality for the perceptual track.
\end{itemize}

The next section describes briefly the method of each team, while in the Appendix A the team members and their affiliations are provided.




\section{Challenge Methods}

\subsection{Islab-zju team}
\label{report:q19911124}

\begin{figure*}[htbp]
	\centering
	\includegraphics[width=1.8\columnwidth]{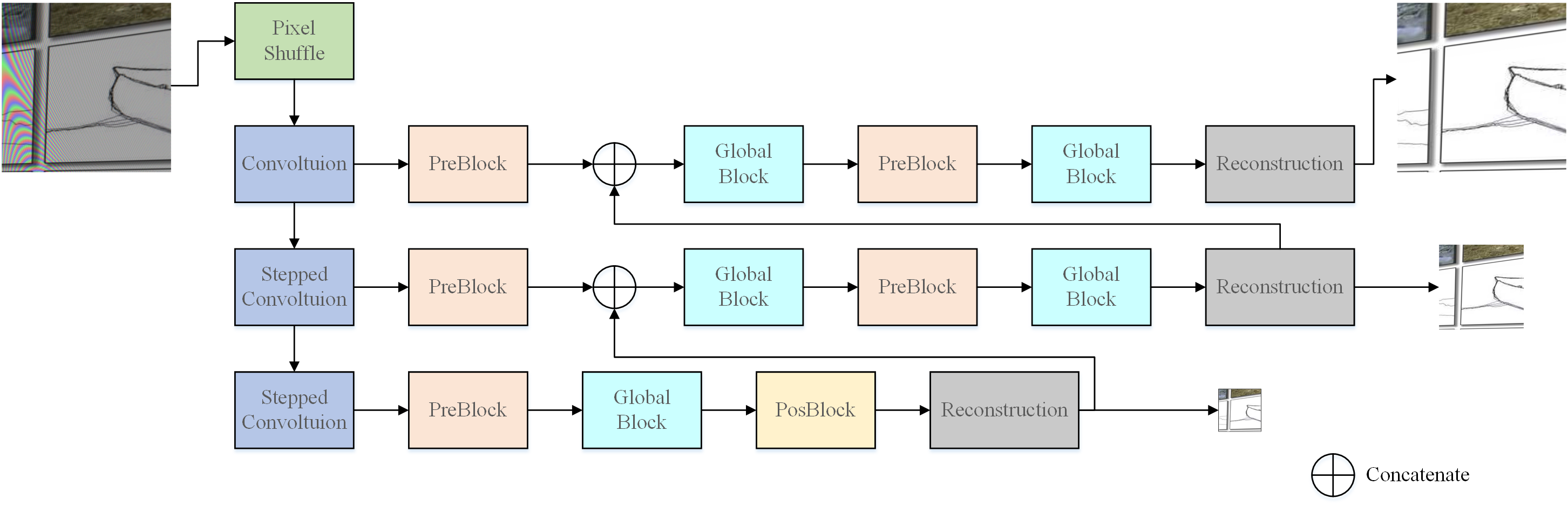}
	\caption{Islab-zju's proposed MBCNN architecture, winner of both tracks of the challenge.}
	\label{fig1}
\end{figure*}

The Islab-zju team proposed a \textit{CNN Based Learnable Multiscale Bandpass Filter method}, dubbed \emph{MBCNN}. Moire artifacts often appear as a multiscale and multi-frequency texture. Based on this observation, and inspired by IDCN~\cite{IDCN}, Islab-zju designed a CNN-based learnable multiscale bandpass filter. The architecture of proposed method is shown in Figure~\ref{fig1}.

MBCNN not only removes the moire artifact, but also enhances the luminance of moire image. 
In practice, the method referenced the architecture of CAS-CNN~\cite{CAS-CNN} and proposes a 3-level multiscale bandpass convolutional neural network, hence the name MBCNN. 
MBCNN first uses pixel shuffle to reduce the input image resolution to half of original image. Subsequent scales are generated using two $2\times2$ stepped convolution layers to start the other two levels. 

\begin{figure}[htbp]
	\centering
	\includegraphics[width=1.0\columnwidth]{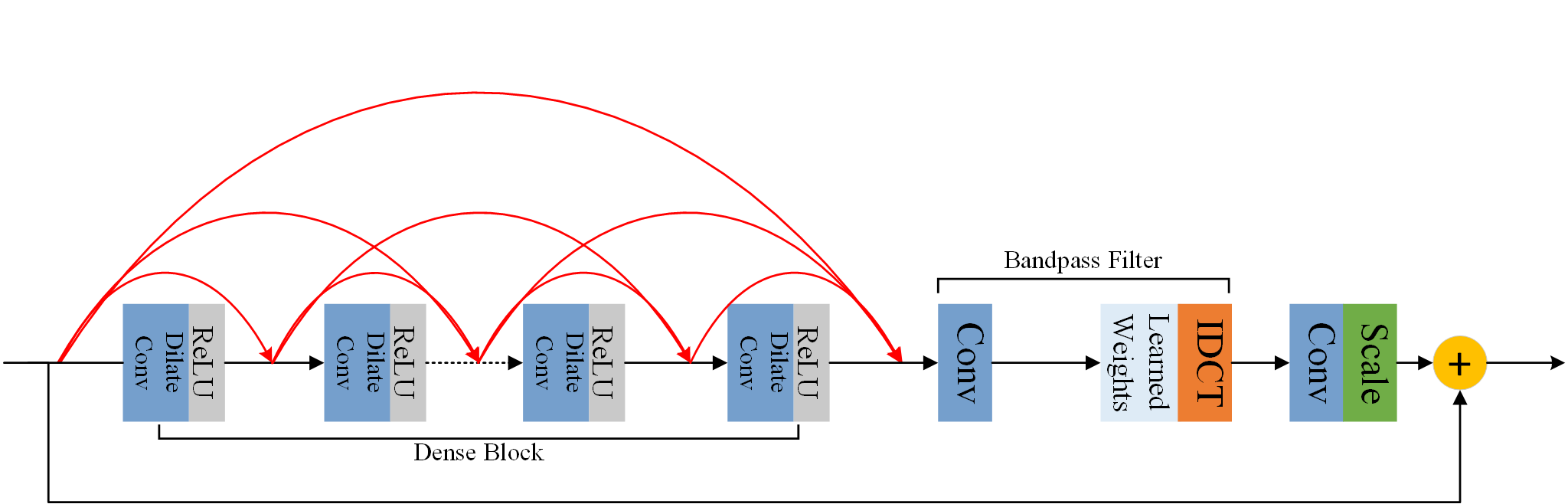}
	\caption{Islab-zju's architecture of $PreBlock$.}
	\label{fig2}
\end{figure}

\begin{figure}[htbp]
	\centering
	\includegraphics[width=1.0\columnwidth]{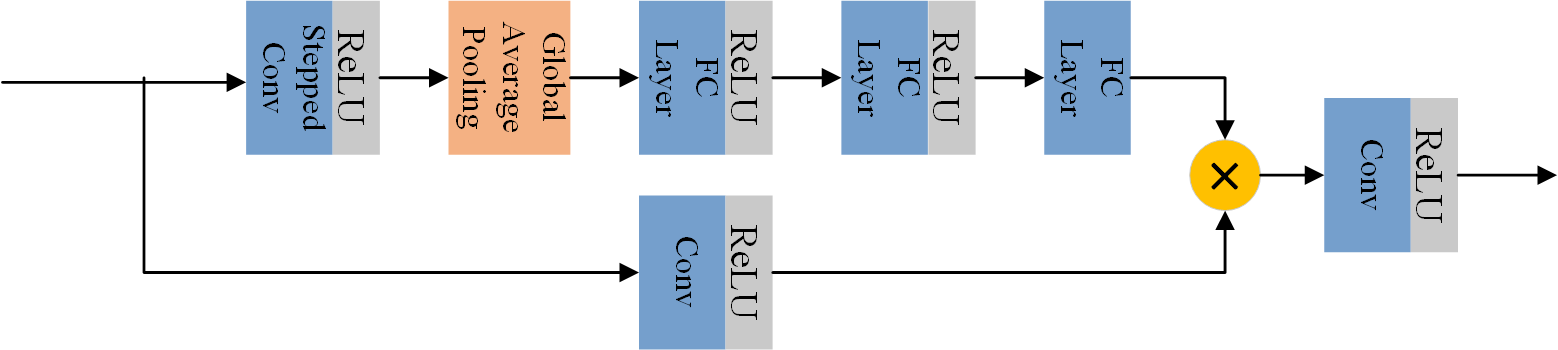}
	\caption{Islab-zju's architecture of global channel attention block.}
	\label{fig3}
\end{figure}

For each level, the processing can be divided into three parts: artifact reduction, tone mapping and image reconstruction. For artifact reduction, MBCNN proposes a residual connection~\cite{ResNet} based block. 
For tone mapping, a dense connection~\cite{DenseNet} based block is used. For image construction, MBCNN uses a convolution layer with a pixel shuffle~\cite{espcn} layer to reconstruct the final output image. The pipeline of each level can be described as removing moire artifact, then enhancing the luminance and finally outputting a clean image in different scales. 
For simplicity, MBCNN denotes the artifact reduction block as $PreBlock$, and tone mapping block as $PosBlock$. 
In a $PreBlock$ (shown in Fig.~\ref{fig2}), MBCNN uses 5 densely connected convolution layers to extract convolutional domain feature. 
Then MBCNN uses a bandpass filter to estimated the convolutional domain loss. 
Finally, a residual connection  connects the input of $PreBlock$ and estimated convolutional domain loss to obtain the clean convolutional domain features. 
In a $PosBlock$, MBCNN uses 5 densely connected convolution layers with a $1\times1$ convolution layer to learn the tone mapping function in the convolutional domain. 
However, only using these two blocks in a level cannot well remove the moire artifact in color-background areas. 
To make a balance for all pixels in an image, MBCNN references the architecture of global block in HDRNet~\cite{hdrnet} and proposes a global channel attention (GCA) block. 
The architecture of GCA block is shown in Figure~\ref{fig3}. MBCNN concatenates a GCA block after each $PreBlock$. The third level contains a $PreBlock$ with a GCA block and a $PosBlock$. The other two levels contain two $PreBlock$s with a GCA block and a $PosBlock$. Specifically, in level 1 and 2, the outputs of level 2 and level 3 will be concatenated to the outputs of the first $PreBlock$s of level 1 and 2.

\begin{figure}[htbp]
	\centering
	\includegraphics[width=0.90\columnwidth]{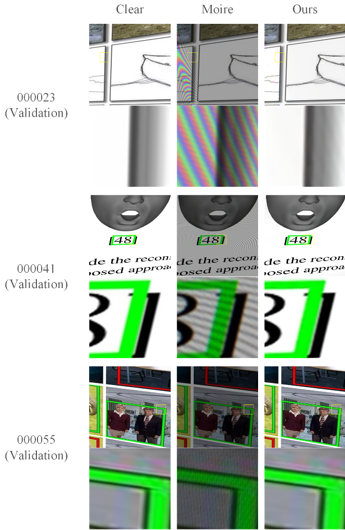}
	\caption{Islab-zju's some cases where MBCNN excels at removing moire artifact.}
	\label{fig4}
\end{figure}

\begin{figure}[htbp]
	\centering
	\includegraphics[width=0.90\columnwidth]{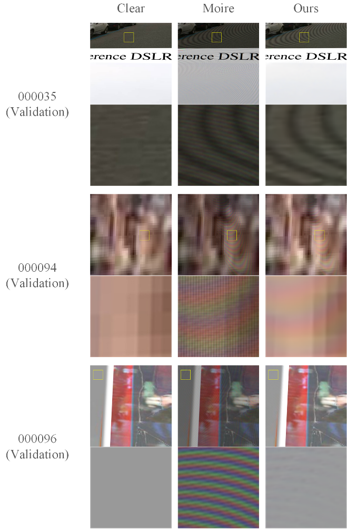}
	\caption{Islab-zju's some cases where MBCNN struggles to remove moire artifact.}
	\label{fig5}
\end{figure}

The loss used to train the network includes L1 loss and L1 Sobel loss. 
They used implicit DCT to estimate frequency loss and introduced learnable weights for each DCT coefficient to construct a bandpass filter.
The method incorporates learnable weights for each DCT coefficient to construct a bandpass filter. 

Generally, as shown in Figure~\ref{fig4}, the proposed method can effectively remove the moire artifact in most scenes. However, when processing some images with extremely complex color texture, or extremely strong moire artifact, MBCNN struggles to completely remove the moire artifact (shown in Figure~\ref{fig5}).

\subsubsection{Implementation details}
The proposed method is developed using the Python language, and can be easily deployed on other platforms. The method obtained all trained model weights, validation and testing results using Windows 10, Python3.7, Keras 2.2.4 with Tensorflow 1.14.0 backend. Training MBCNN on a 2080Ti GPU took roughly 2 days. The training images are cropped into $128\times128$ patches for training. The batchsize is set to 16 and 8000 batches construct a epoch. Adam~\cite{Adam} was used with default settings as the optimizer. The learning rate was initialized to $10^{-4}$ and halved by every 10 epochs. At inference, for processing a $1024\times1024$ color image, MBCNN roughly consumed 10GB memory and took 0.25s on a 2080Ti GPU.

\subsection{MoePhoto team}
\label{report:opteroncx}


\begin{figure*}[th]
\begin{center}
\includegraphics[width=0.8\linewidth]{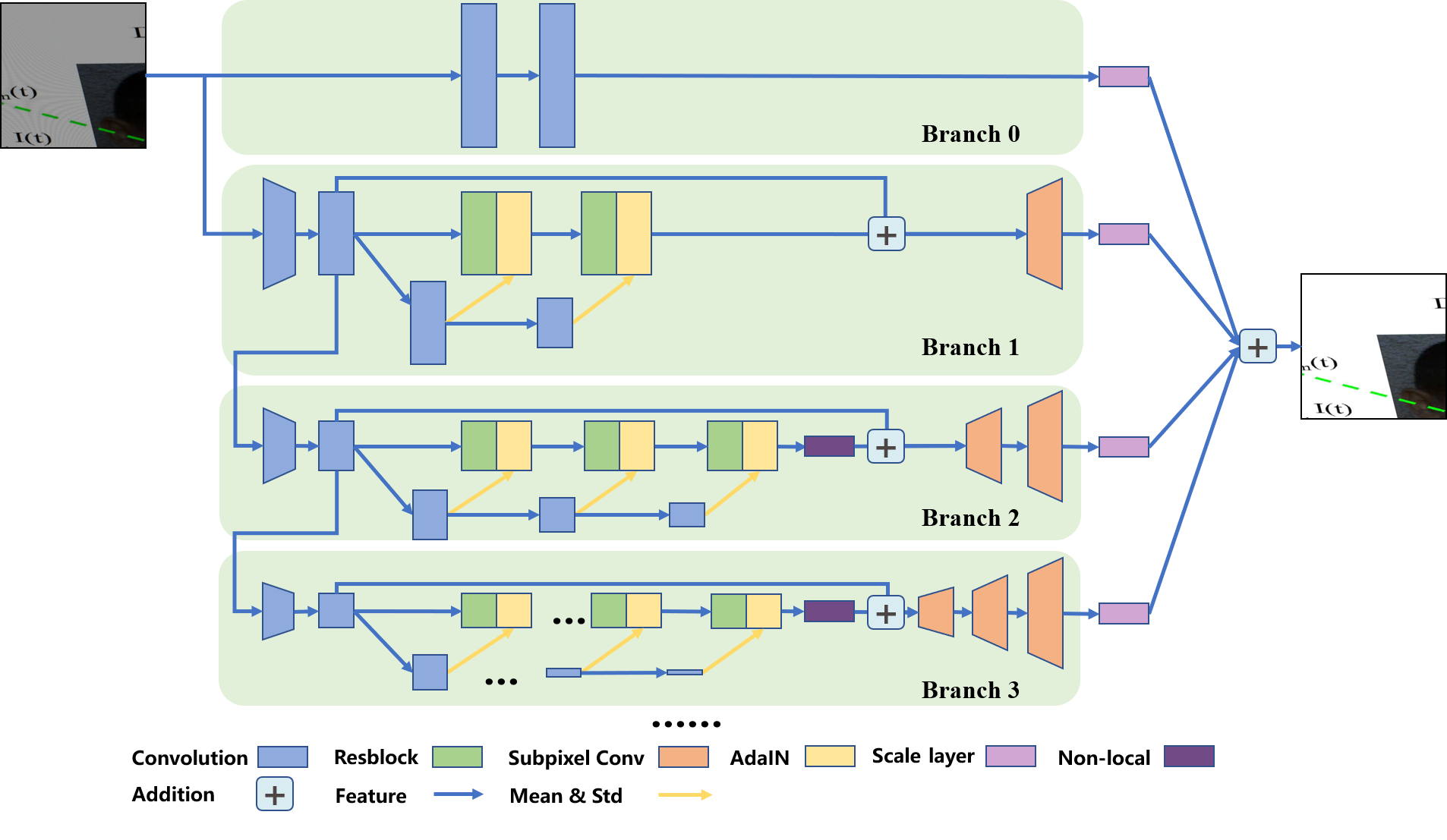}
\end{center}
   \caption{MoePhoto's Overall structure of MDDM. }
\label{fig:short}
\end{figure*}

The MoePhoto team proposed the method of \textit{Multi Scale Dynamic Feature Encoding Network for Image Demoireing}, namely MDDM~\cite{cheng2019multi}, shown in Figure~\ref{fig:short}. This method is inspired by \cite{liu2018demoir, gao2019moire, sun2018moire}

In order to remove the moire pattern, MDDM constructs an image feature pyramid, encodes image features at different spatial resolutions, and obtains image feature representations of different frequency bands. 
MDDM incorporates a multi-scale residual network with multiple resolution branches. These branches learn the nonlinear mapping on the original resolution and the $2\times$, $4\times$, $8\times$, $16\times$ and $32\times$ downsampled features. Then, the features are upsampled to the original resolution with sub-pixel convolution at the end of each branch. 
The network automatically learns the weight of each branch and sums the results to obtain the final output. 

MDDM also proposes a dynamic feature encoding (DFE) method. At each downsampling branch, the method adds an extra lightweight branch. The number of convolutional layers in this extra branch is equal to the number of residual blocks in the backbone branch. Each dynamic feature coding branch learns the characteristics of the global residual at different scales and affects the main branch by adaptive instance normalization.


\subsubsection{Implementation details}

The total method complexity is reported to be 472.38 GFLOPs($1024\times1024\times3$ input) and 8.01M parameters. MDDM pretrained VGG19 in a GAN version during in the training stage for computing the perceptual loss.
Pytorch 1.2 was used to build the MDDM model and training took place on an NVIDIA Titan V GPU with CUDA10.0. The Adam optimizer was used when training the model. 
The authors first trained a basic model which only contains three branches for feature decomposition and demoireing. 
The initial learning rate was 1e-4, and the learning rate was reduced by 10 times for every 30 epochs. 
Then they further increased the network size, gradually increased to 4 branches, 5 branches and 6 branches and fine tuned these models on the previous basis. 
On the fine tuning stage, the learning rate was set to 1e-5 and reduced by 10 times for every 50 epochs.
At inference, the average running time of the proposed method is 0.15s per image with NVIDIA Titan V GPU for the fidelity track.
The method uses an L1 Charbonnier loss as the loss function in the \textbf{Fidelity} track, and is implemented as GAN version for perceptual optimization in the \textbf{Perceptual} track.
The training time is 5 days for the Basic model (3 branches) and 3 days per stage for finetuning, the runtime is 0.11s per image for the perceptual track.


\subsection{XMU-VIPLab team}
\label{report:xdhm2017}

XMU-VIPLab team proposed \textit{Global-local Fusion Network for Single Image Demoireing}, shown in Figure~\ref{fig:short}. The proposed method first trains a global and a local demoireing network, respectively, and then uses a fusion branch to fuse the results of the two networks to obtain the final demoireing result. The local network uses an asymmetric encode-decode architecture, using the Residual in Residual blocks as the basic block, and the input is a $512 \times 512$-size image cropped
from the training images. The global network uses a standard U-Net architecture, and the input is scaled by 4. The fusion network only contains 4 residual modules. When fusing, the result of the global network needs to be amplified by 4 times.

\begin{figure*}[th]
\begin{center}
\includegraphics[width=0.9\linewidth]{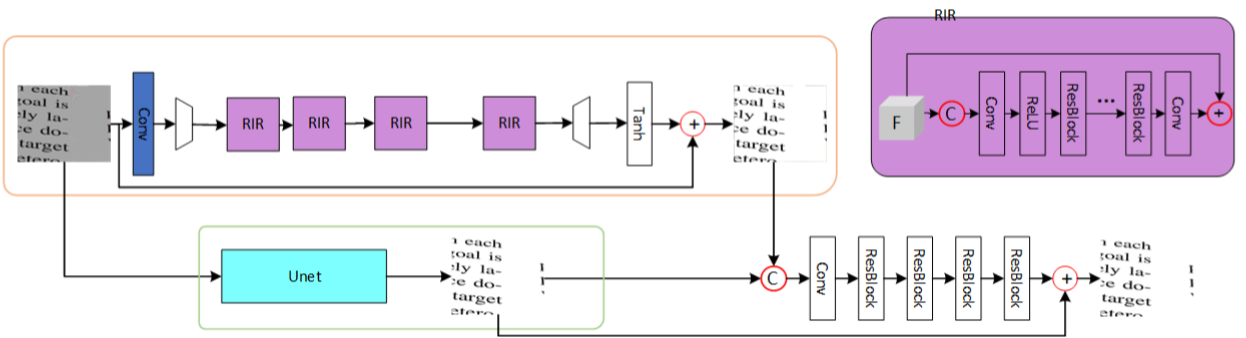}
\end{center}
   \caption{XMU-VIPLab's overall structure of Global-local Fusion Network for Single Image Demoireing. }
\label{fig:short}
\end{figure*}

\subsection{KU-CVIP team}
\label{report:GibsonGirl}

\begin{figure*}[ht]
        \centering
        \includegraphics[width=.9\linewidth]{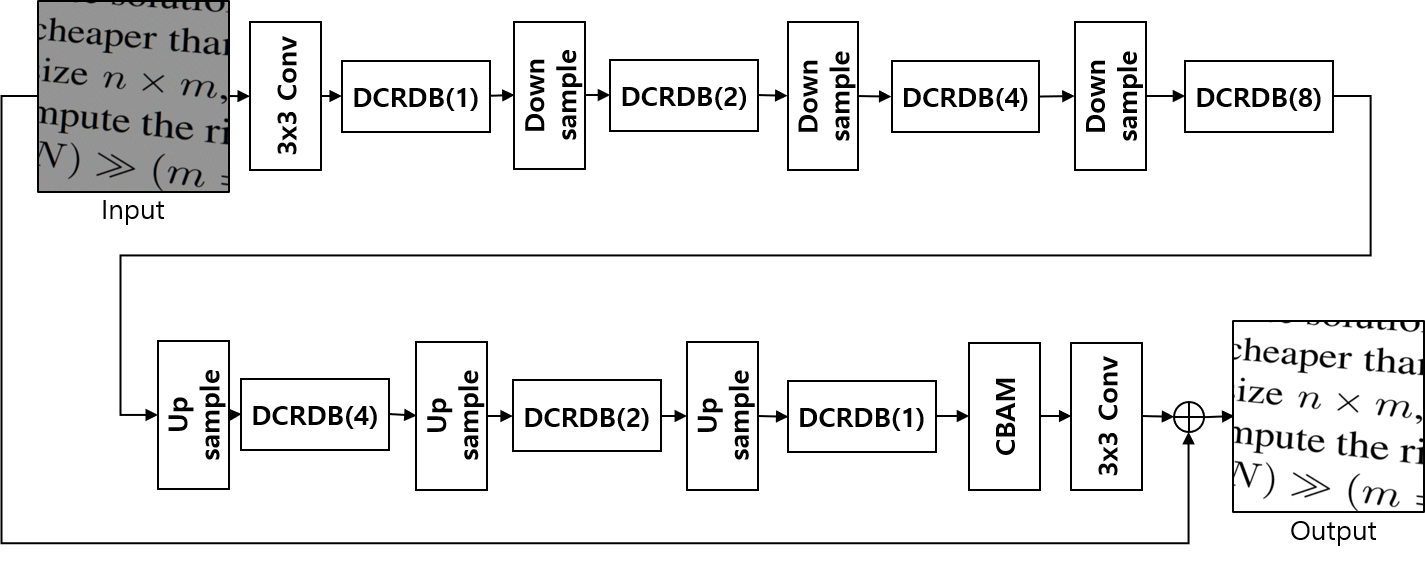}
        \caption{KU-CVIP's Densely-Connected Residual Dense Network Structure.}
        \label{fig:my_label1}
    \end{figure*}

\begin{figure}[h]
        \centering
        \includegraphics[width=0.9\linewidth]{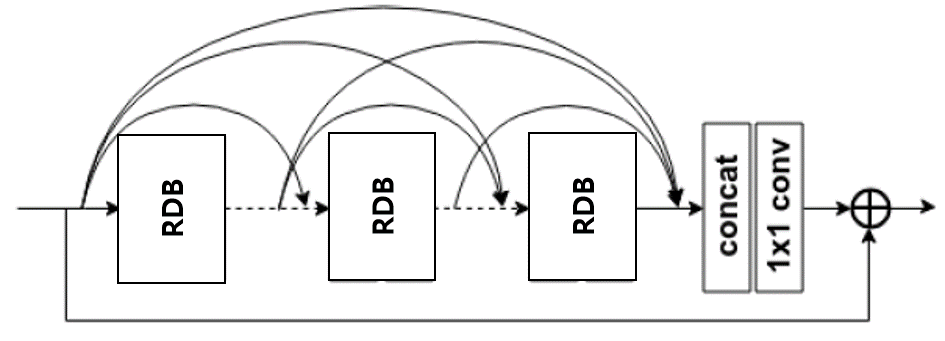}
        \caption{KU-CVIP's DCRDB Structure}
        \label{fig:my_label2}
    \end{figure}

\begin{figure}[h]
        \centering
        \includegraphics[width=0.9\linewidth]{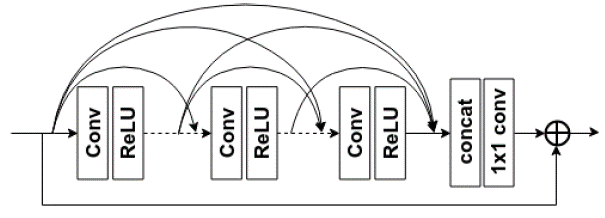}
        \caption{KU-CVIP's RDB Structure}
        \label{fig:my_label3}
    \end{figure}

The KU-CVIP team proposed the \textit{Densely-Connected Residual Dense Network}, inspired by~\cite{RDN, kim2019grdn}
The proposed network is based on Residual Dense Block (RDB). Originally, RDB was first proposed for super-resolution in~\cite{RDN}. 
Afterwards, the authors of \cite{kim2019grdn} showed that the grouped RDBs are effective for removing noise in an image. 
The KU-CVIP team's approach connects RDBs densely and adds skip connections in each DCRDB unit (see Figures~\ref{fig:my_label1},~\ref{fig:my_label2}, and \ref{fig:my_label3}. 
The method downsamples and upsamples the feature map using 4$\times$4 convolution layers with stride 2 to extract multi-scale features and make the network deeper. 
Starting with 1 RDB in a DCRDB unit, the method increases/decreases the number of RDBs in DCRDB by a factor of 2 whenever downsampling/upsampling the feature map. 
At the end of the network, the method applies a CBAM~\cite{woo2018cbam} module and 3$\times$3 convolution layer to make the input and output image have same number of channels.



\subsubsection{Implementation details}

In total, there are 184 convolution layers.
For training, experiments were conducted on a Titan X Pascal. 
As the limit of the GPU memory, the method randomly cropped a 128$\times$128 image patch from original image and was trained using a batch size of 8. 
The network removes all batch normalization layers. The objective function is an L1 loss optimized  using Adam with beta1 = 0.9 and beta2 = 0.999. Initial learning rate was 0.0001 and was decreased by half at every 100 epochs. The model was trained for 54 hours.

\subsection{IAIR team}
\label{report:wangjie}

\begin{figure*}[ht]
    \centering
    \includegraphics[width=0.9\linewidth]{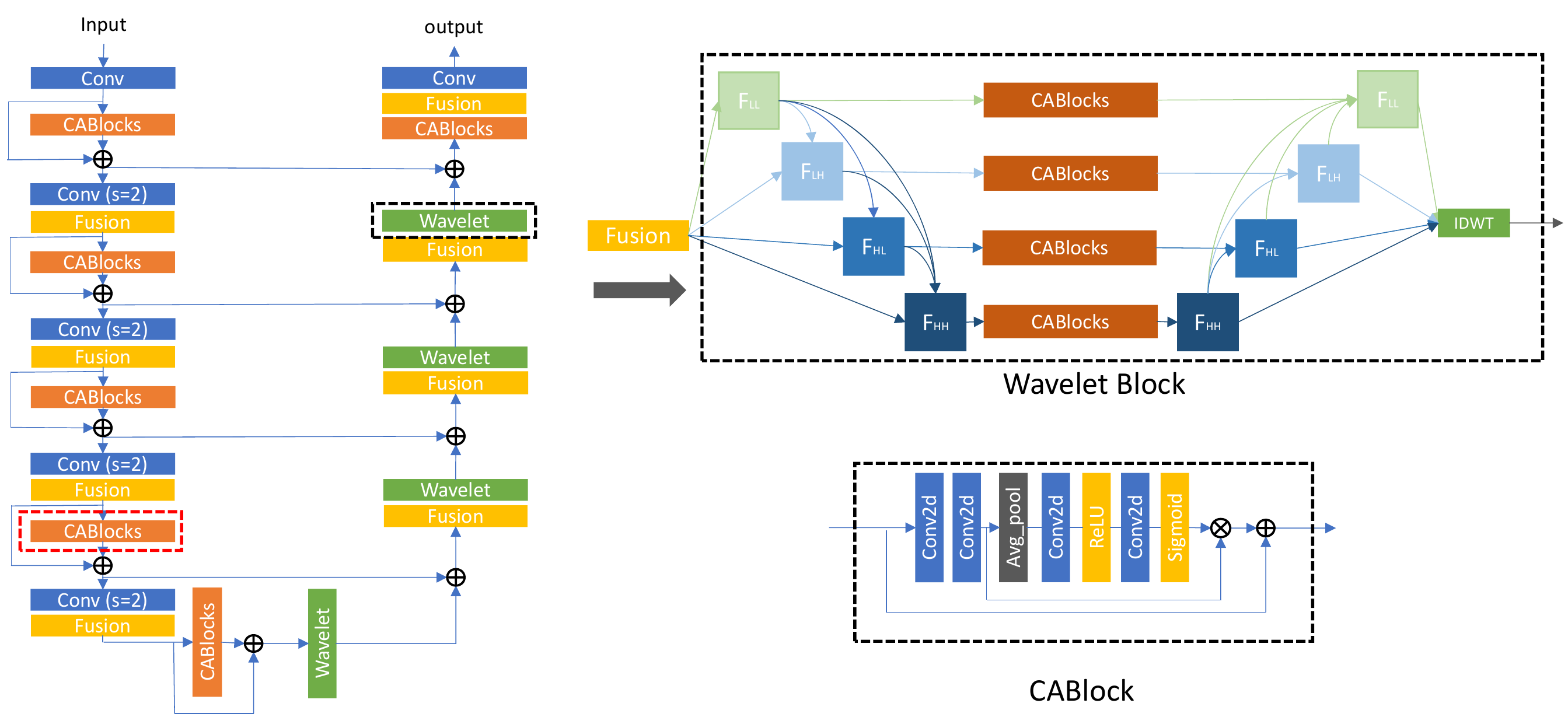}

    \caption{IAIR's architecture of the proposed model.}
    \label{fig:1}
\end{figure*}

The IAIR team proposed \textit{Wavelet-Based Multi-Scale Network}, namely WMSN, for Image Demoireing, see Figure~\ref{fig:1}.

This network is based on the U-Net~\cite{UNet} architecture, in which the clique wavelet block~\cite{nips18} was used in the decoder part to restore the degraded features 
and upsample the restoration features.
As shown in Figure~\ref{fig:1}, the proposed network contains four strided convolutional layers and four wavelet blocks. 
The first four CABlocks in encoder part consists of three residual channel attention blocks~(RCAB) by~\cite{rcan}, and 18 RCABs are used in the last CABlocks. 
The clique wavelet blocks are used to remove the moire artifacts from the features in the wavelet domain and upsample the restoration features. 
The wavelet block first uses four convolutional layers to disentangle the features into four wavelet domains denoted by $F_{LL}, F_{HL}, F_{LH}$, and $F_{HH}$.
Then, each branch is fed into four CABlocks and one convolutional layer to restore the features.
Finally, the method applies a Haar inverse discrete wavelet transform (IDWT) to combine the features from different wavelet domains and increase the resolution.
The filter size is set as 11 $\times$ 11 pixels in the first convolutional layer in the encoder part and 3 $\times$ 3 in all other convolutional layers.

\begin{figure}[tb]
    \small
    \centering
    \begin{tabular}{cccc}
      \hspace{-3mm}
       \includegraphics[width=0.44\linewidth]{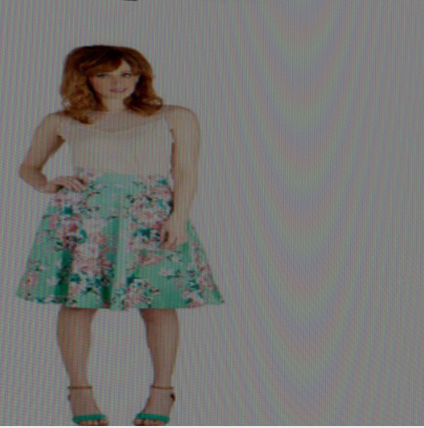} & \hspace{-4mm}
        \includegraphics[width=0.44\linewidth]{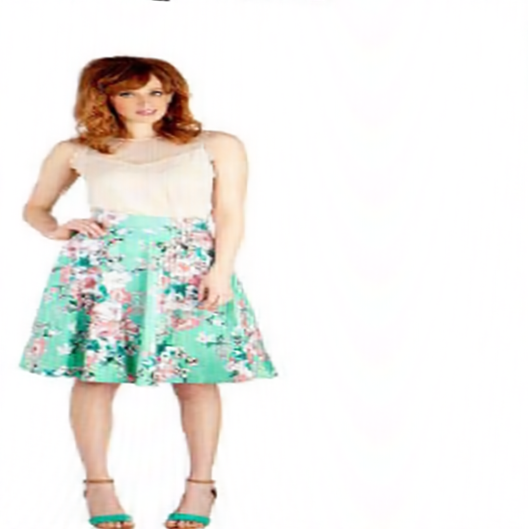} & \hspace{-4mm}
      \vspace{-0.5mm}
      \\
      \hspace{-3mm} 
        (a) Moire input & \hspace{-4mm}
      (b) Output & \hspace{-4mm}
  
      \\
      \hspace{-3mm}
        \includegraphics[width=0.44\linewidth]{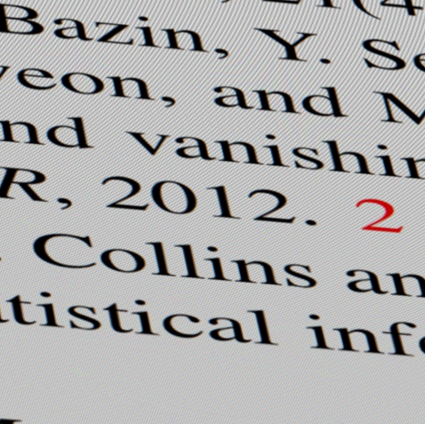} & \hspace{-4mm}
        \includegraphics[width=0.44\linewidth]{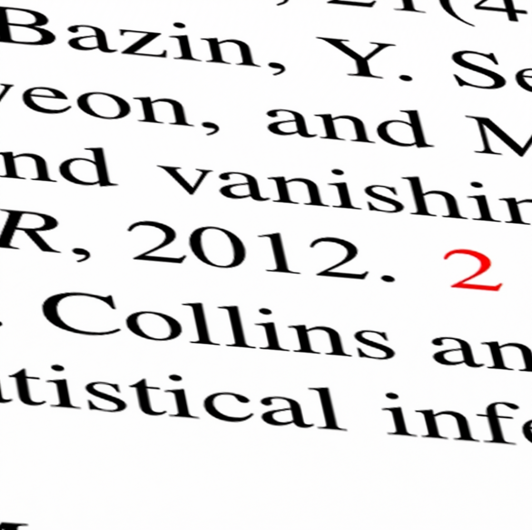} & \hspace{-4mm}
        \vspace{-0.5mm}
      \\
      \hspace{-3mm} 
        (c) Moire input & \hspace{-4mm}
        (d) Output & \hspace{-4mm} 
  
    \end{tabular}
    \caption{IAIR's visual results on the test set.}
    \label{fig:visual_results_GOPRO}
\end{figure}

\subsubsection{Implementation details}

The network has 38.2 million parameters and $8.4 * 10^{10}$ FLOPS in inference for an $1024 \times 1024$ input.
The authors trained a simple classification network to divide the training, validation, and test set into two parts: one for text images and one for other images.
Then the approach was trained twice; separately for these two classes.
The model was trained on the training set with augmentation as in~\cite{GFN}. 
During the training phase, patches with a size of $64 \times 64$ were cropped from training images as inputs and the batch size is set to 16. 
The network was trained with L2 loss and perceptual loss~\cite{vggloss} using the Adam solver with $\beta_1 = 0.9$ and $\beta_2 = 0.999$. 
An initial learning rate of 1e-4 was used, with a decay rate of 0.1 every 50 epochs. %
The network was trained from scratch for 80 epochs. 
All the training and testing processes are conducted on an NVIDIA 1080Ti GPU.
In the testing phase, first the pre-trained classification network is used to divide the test images into text images and other images.
Then the method uses the text model to restore the text images and the other model to restore non-text images.
To analyze the advantages of the proposed solution, the authors first train a baseline model by replacing the wavelet blocks with deconvolutional layers.
The baseline model is trained on the same settings with the proposed model and achieves 36.9 dB PSNR on a validation set. 
After adding the wavelet blocks, the proposed model obtain 1.9 dB PSNR gain on the validation set (38.8 dB PSNR). 
For training the image classification model, the authors manually select 500 textual images and 500 other images from training set. The model was trained for 5 days.

\subsection{PCALab team}
\label{report:hanzy}

\begin{figure}[ht]
    \centering
    \includegraphics[width=\columnwidth]{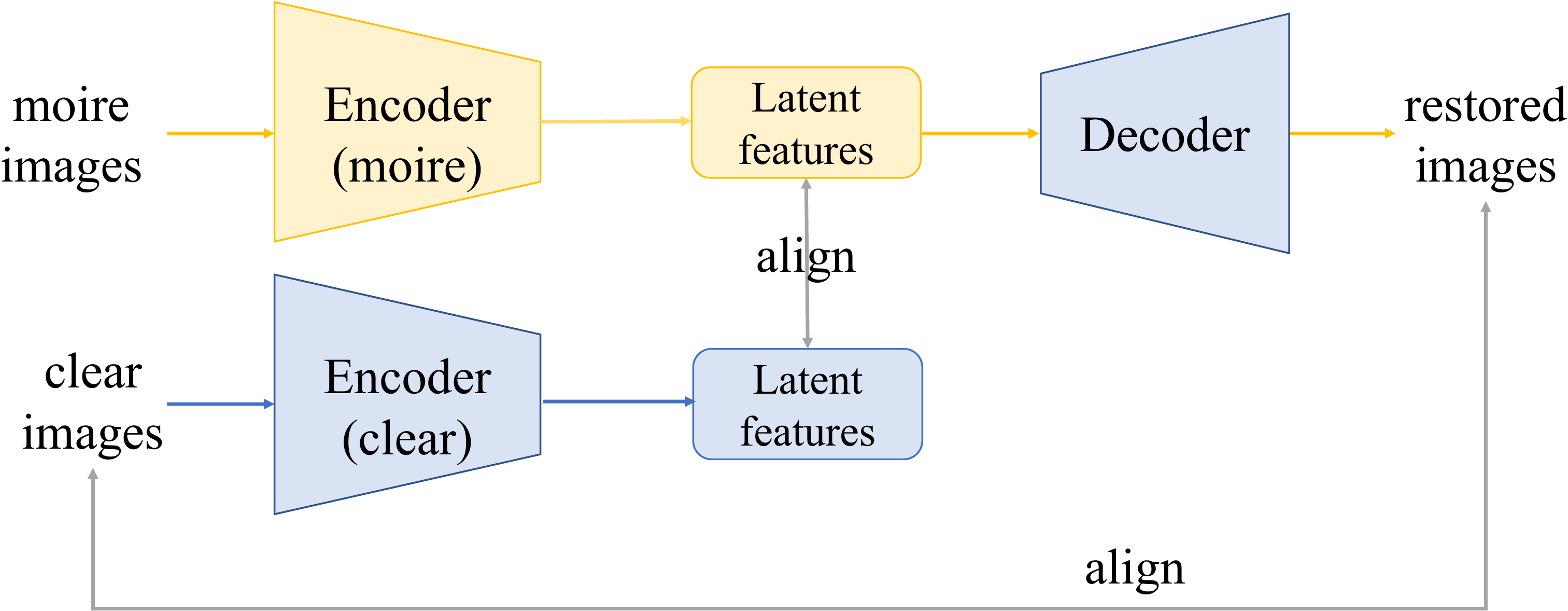}
     \caption{PCALab's Framework of Domain adaptation for image demoireing.}
     \label{fig:hanzy}
  \end{figure}

The PCALab team proposed \textit{Domain adaptation for image demoireing}, as shown in  Figure~\ref{fig:hanzy}. The method is inspired by~\cite{tzeng2017adversarial, mechrez2018contextual}.
The method proposes to remove moir\'{e} pattern through a domain adaptation method, in which the method assumes that the clear images come from clear domain and the images with moir\'{e} pattern are from a moir\'{e} domain and the two domains share the same latent space. 
The method incorporates an Auto Encoder (AE) framework, where the encoder is designed to extract the common latent features shared by two domains and the decoder restores the images from corresponding latent features. 
The method first learns an Auto Encoder (AE) on clear domain. Then another encoder for moir\'{e} domain is introduced to extract the latent features from the moir\'{e} images and the latent features of the moir\'{e} images are supposed to be the same as the latent features of the corresponding clear images, which is extracted by the encoder for clear domain. %
When the method learns the encoder for the moir\'{e} domain, the latent features of moir\'{e} images extracted by the encoder for the moir\'{e} domain are just the same with the latent features of clear images extracted by the encoder for the clear domain. 
Then given a moir\'{e} images, the common decoder can synthesize corresponding clear images conditioned on the latent features extracted by the encoder for the moir\'{e} domain.

In the \textbf{fidelity} track, the method aligns the latent features of moir\'{e} images and clear images by minimizing their L1 distance. Also included is the contextual loss to further make the latent features of moir\'{e} images close to the the latent features of clear images.
The method also learns the encoder for the moir\'{e} domain by minimizing the L1 distance between the images generated from latent features of moir\'{e} images and the corresponding clear images. 
In the \textbf{perceptual} track, the contextual loss is removed.

\subsubsection{Implementation details}

The models had 169.72 GFLOPs for the \textbf{fidelity} track and 95.26GFLOPs for the \textbf{perceptual} track for an $1024\times 1024\times 3$ input. There are 4.98M parameters for the \textbf{fidelity} track and 2.13M parameters for the \textbf{perceptual} track.
The encoder for the clear domain and the decoder are pretrained in the first stage and then fixed when learning the encoder for the moir\'{e} domain.
%
The networks were trained using the Adam optimizer when training the model with learning rate 1e-4. First an encoder for the clear domain was trained, and the decoder on the clear image set. Then an encoder for the moir\'{e} domain based on the paired samples containing moir\'{e} images with corresponding clear images was trained.
%
%
%
In the \textbf{fidelity} track, the decoder contains 6 hidden layers and both of the encoder contains 12 layers. In the \textbf{perceptual} track, all of the three models contain 6 hidden layers.  
The training cost about 3 days. Test time: less than 0.02s per image.

\subsection{Neuro team}
\label{report:souryadipta}

The Neuro team proposed the \textit{Demoireing High Resolution Images Using Conditional GANs}.
This method models the problem as a image translation problem and uses cGANs to reconstruct the demoired image. 
The frequency spectrum of moire images have some high frequency components missing due to aliasing. GANs are used widely used to reconstruct an image, including different types of signals. Therefore, reconstructing those high frequency components from rest of the frequency spectrum of a image is the main goal for this solution. 
Here, the method employs a Global Generator instead of local Generator to emphasize global features more to increase fidelity and the perceptual quality of the restored image.

As described in the work of Liu~\etal~\cite{liu2018demoir}, moire effect is heavily dependent on scale of the image.  Therefore, this method performs the image translation in high resolution, which will ultimately take less time as in a lower scale.  Therefore the method needs to do upsampling which is a computationally heavy task. In this cGAN, the method uses a multi-scale discriminator to solve this problem. However, the authors noted that training both networks is a difficult task because of instability problems. More detail about the architecture and training of this cGAN can be found in Wang~\etal~\cite{wang2018high}.

\subsubsection{Implementation details}

The model consists of several 2D convolutional layer, Resnet Blocks, instance normalization blocks etc. 
%
%
The model is trained with batch size of 1 with Adam optimizer with initial learning rate 0.0002. In the generator, number of downsampling layers used is 4 and number of residual block is 9. In the input layer, size is 1024 x 1024.
%
The approach only used the Global Generator to generate the image with batch size of 1.
%
The approach basically used Wang~\etal~\cite{wang2018high}'s work from NVIDIA with their implementation of \href{https://github.com/NVIDIA/pix2pixHD}{pix2pixHD}.
The authors came up with the idea to use this publication as theoretically this work solves nearly all required challenges of this demoireing problem efficiently.
%
%
%
%
%
The training time is 4-5 days.
%

\subsection{IPCV\_IITM team}
\label{report:kuldeeppurohit3}

\begin{figure*}[th]
\centering
\includegraphics[scale=0.52]{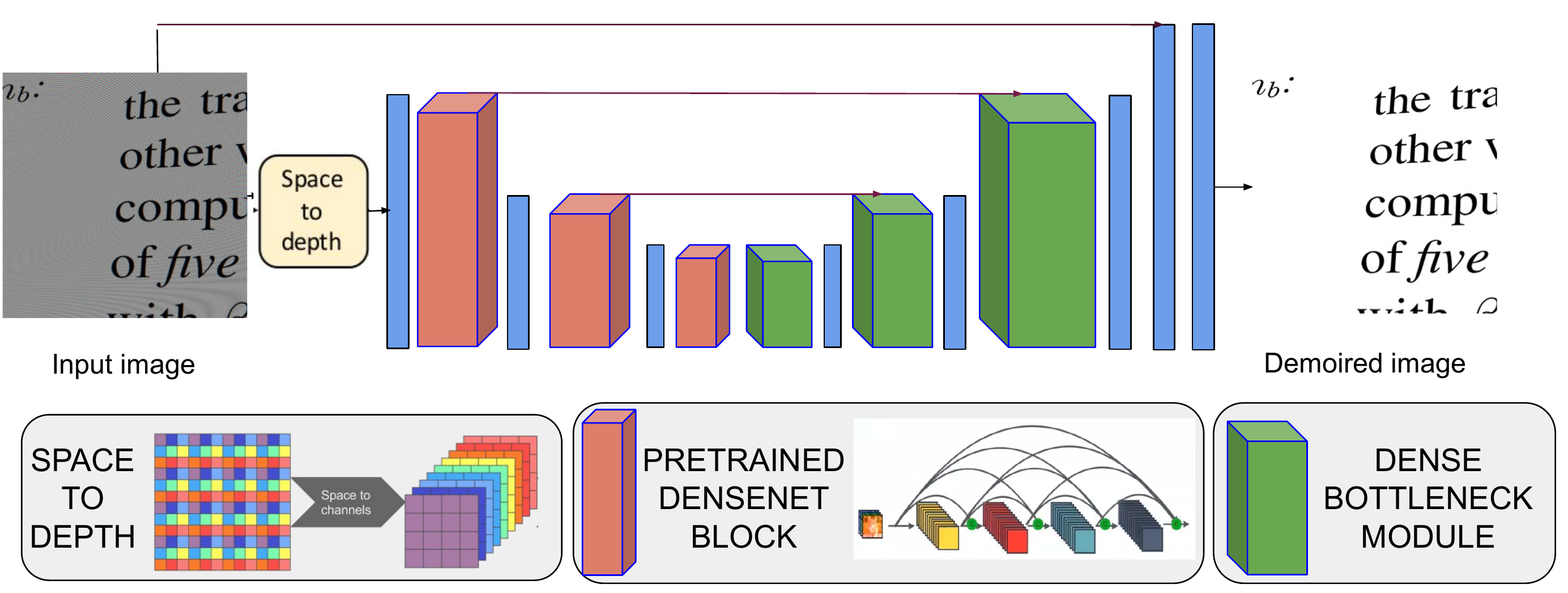} \\
\caption{IPCV\_IITM's Structural details of our Sub-pixel Dense U-net.}
\label{fig:diagram}
\end{figure*} 

The IPCV\_IITM team proposed the \textit{Sub-pixel Dense U-net for Image Demoireing}, as shown in  Figure~\ref{fig:diagram}.
The proposed network utilizes a similar backbone structure as the restoration work of~\cite{2019arXiv190311394P}.
It consists of a deep dense-residual encoder-decoder structure with subpixel convolution and multi-level pyramid pooling module for efficiently estimating the demoireing image. 
The resolution of input image is brought down by a factor of 2 using pixel-shuffling before being fed to the encoder, which increases the receptive field of the network and reduces computational footprint. The encoder is made of densely connected
modules. It helps to address the issue of vanishing gradients and feature
propagation while substantially reducing the model complexity. The weights
are initialized using pre-trained Dense-121 network~\cite{DenseNet}. Each layer in a block
receives feature maps from all earlier layers, which strengthens the information flow during forward and backward pass making training deeper
networks easier.
The final output is the residual between the ground-truth colored image
and the input image.

The Decoder accepts the features estimated by the encoder at various levels and processes them using residual blocks before increasing their spatial
resolution through bi-linear up-sampling and convolution. The intermediate features with higher spatial resolution in the decoder are concatenated
with the corresponding-sized encoder features. Finally, the decoder output
is enhanced through multi-scale context aggregation through pooling and
upsampling at 4 scales, before being fed to the final layer. To match the
features to the input image resolution, the method employs a subpixel-convolution
block to upsample them by a factor of 2. The inference
time of the network is optimized by performing computationally intensive operations
on features at lower spatial resolution. This also reduces memory footprint
while increasing the receptive field. The method makes the downsample-upsample
operation end-to-end trainable which gives better performance compared
to bilinear operations (used in \cite{ancuti2018ntire}). The upsampled features are passed through a convolutional layer to construct the demoireing output.

                                  
\subsubsection{Implementation details}


The total method complexity was $\approx$ 10 million parameters.
The first three blocks of dense-net 121 was trained on Imagenet.
In order to maximize the potential performance of the model, a self-ensemble strategy was adopted. During the test time, the input image was flipped and rotated to generate seven augmented inputs. With those augmented images, the method generates corresponding demoireing images using the net-
works. Then inverse transform is applied to each of the output images. Finally,
they average the transformed outputs all together to make the self-ensemble result.

\section{Conclusion}

This paper reviews the first-ever image demoireing challenge. First, the paper describes the challenge including the new dataset, LCDMoire that was created for participants to develop their demoire methods as well as evaluate proposed solutions. The challenge consisted of two tracks (fidelity and perceptual) that considered different image quality assessment techniques to characterize the demoire approaches submitted by participants.  The tracks had 60 and 39 registered participants, respectively, and a total of eight teams competed in the final testing phase. The paper summarizes the results of the two tracks, and then describes each of the approaches that competed in the challenge. The entries span the current the state-of-the-art in the image demoireing problem. The Islab-zju team's approach was the leading demoire method, winning both tracks. We hope this challenge and its results will inspire additional work in the demoire problem, which is becoming an increasingly important image quality challenge.

\section*{Acknowledgments}
We thank the AIM 2019 sponsors.

\appendix
\section{Teams and affiliations}
\label{report:teams}

\subsection{AIM 2019 team}
\begin{itemize}
\item \textbf{Title:} 
AIM 2019 Challenge on Image Demoireing
\item \textbf{Members:} \\
Shanxin Yuan$^1$ (\textcolor{cyan}{shanxin.yuan@huawei.com}) \\
Radu Timofte$^2$  (\textcolor{cyan}{radu.timofte@vision.ee.ethz.ch})\\
Greg Slabaugh$^1$  (\textcolor{cyan}{gregory.slabaugh@huawei.com})\\
Ale\v{s} Leonardis$^1$   (\textcolor{cyan}{ales.leonardis@huawei.com})
\item \textbf{Affiliations:} \\
$^1$Huawei Noah's Ark Lab \hspace*{4mm}\\
$^2$ETH Z\"{u}rich, Switzerland
\end{itemize}

\subsection{IsLab-zju team}
\begin{itemize}
\item \textbf{Title:} 
CNN Based Learnable Multiscale Bandpass Filter for Image Demoireing
\item \textbf{Members:} \\
Bolun Zheng (\textcolor{cyan}{zhengbolun1024@163.com}), Xin Ye, Xiang Tian, Yaowu Chen
\item \textbf{Affiliation:} 
Zhejiang University, China
\end{itemize}

\subsection{MoePhoto team}
\begin{itemize}
\item \textbf{Title:} 
Multi scale dynamic feature encoding network for
image demoireing
\item \textbf{Members:} \\
Xi Cheng (\textcolor{cyan}{chengx@njust.edu.cn}), Zhenyong Fu, Jian Yang
\item \textbf{Affiliation:} 
Nanjing University of Science and Technology, China
\end{itemize}

\subsection{XMU-VIPLab team}
\begin{itemize}
\item \textbf{Title:} 
Densely-Connected Residual Dense Network
\item \textbf{Members:} \\
Ming Hong (\textcolor{cyan}{1184307675@qq.com}), Wenying Lin, Wenjin Yang, Yanyun Qu
\item \textbf{Affiliation:}
Xiamen university, China

\end{itemize}

\subsection{KU\_CVIP team}
\begin{itemize}
\item \textbf{Title:} 
Densely-Connected Residual Dense Network
\item \textbf{Members:}\\ 
Hong-Kyu Shin (\textcolor{cyan}{hgshin@dali.korea.ac.kr}), Joon-Yeon Kim, Sung-Jea Ko
\item \textbf{Affiliation:}
Korea University, Korea
\item \textbf{Webpage:}
\url{https://dali.korea.ac.kr}

\end{itemize}

\subsection{IAIR team}
\begin{itemize}
\item \textbf{Title:} 
Wavelet-Based Multi-Scale Network for Image Demoireing
\item \textbf{Members:} \\
Hang Dong (\textcolor{cyan}{dhunter1230@gmail.com}), Yu Guo, Jie Wang, Xuan Ding  
\item \textbf{Affiliation:} 
Xian Jiaotong University, China.
\end{itemize}

\subsection{PCALab team}

\begin{itemize}
\item \textbf{Title:} Domain adaptation for image demoireing
\item \textbf{Members:}\\
Zongyan Han (\textcolor{cyan}{hanzy@njust.edu.cn}), Xi Cheng, Zhenyong Fu, Jian Yang
\item \textbf{Affiliation:} Nanjing University of Science and Technology
\end{itemize}

\subsection{Neuro team}
\begin{itemize}               
\item \textbf{Title:} Demoireing High Resolution Images Using Conditional GANs
\item \textbf{Members:}\\ Sourya Dipta Das (\textcolor{cyan}{dipta.juetce@gmail.com})
\item \textbf{Affiliation:} Jadavpur University,Kolkata,India
\end{itemize}

\subsection{IPCV\_IITM team}

\begin{itemize}               
\item \textbf{Title:} Sub-pixel Dense U-net for Image Demoireing
\item \textbf{Members:} \\
Kuldeep Purohit (\textcolor{cyan}{kuldeeppurohit3@gmail.com}), Praveen Kandula, Maitreya Suin, A.N. Rajagoapalan
\item \textbf{Affiliation:} Indian Institute of Technology Madras, India.
\item \textbf{Team website URL:}
\url{http://www.ee.iitm.ac.in/ipcvlab/}
\end{itemize}

{\small
\bibliographystyle{ieee_fullname}
\bibliography{egbib}
}

\end{document}